\documentclass[conference]{IEEEtran}
\IEEEoverridecommandlockouts
\usepackage{adjustbox}
\usepackage{cite}
\usepackage{amsmath,amssymb,amsfonts}
\usepackage{amsthm}
\usepackage{algorithmic}
\usepackage{graphicx}
\usepackage{textcomp}
\usepackage{xcolor}
\usepackage{algorithm}
\usepackage{array}
\usepackage[caption=false,font=normalsize,labelfont=sf,textfont=sf]{subfig}
\usepackage{textcomp}
\usepackage{stfloats}
\usepackage{url}
\usepackage{verbatim}
\def\BibTeX{{\rm B\kern-.05em{\sc i\kern-.025em b}\kern-.08em
    T\kern-.1667em\lower.7ex\hbox{E}\kern-.125emX}}
\begin{document}

\title{
An Analysis on Stabilizability and Reliability Relationship in Wireless Networked Control Systems
}
\author{\IEEEauthorblockN{Zeinab Askari Donbeh, Mehdi Rasti, Shiva Kazemi Taskooh, and Mehdi Monemi}
\IEEEauthorblockA{Centre for Wireless Communications (CWC),
University of Oulu, Oulu, Finland\\
Emails: \{zeinab.askaridonbeh, mehdi.rasti, shiva.kazemitaskooh, mehdi.monemi\}@oulu.fi}
}
\maketitle

\begin{abstract}
The stabilizability of wireless networked control systems (WNCSs) is a deterministic binary valued parameter proven to hold if the communication data rate is higher than the sum of the logarithm of unstable eigenvalues of the open-loop control system. In this analysis, it is assumed that the communication system provides a fixed deterministic transmission rate between the sensors and controllers. Due to the stochastic parameters of communication channels, such as small-scale fading, the instantaneous rate is an intrinsically stochastic parameter. In this sense, it is a common practice in the literature to use the deterministic ergodic rate in analyzing the asymptotic stabilizability. Theoretically, there exists no work in the literature investigating how the ergodic rate can be incorporated into the analysis of asymptotic stabilizability. Considering the stochastic nature of channel parameters, we introduce the concept of {\it probability of stabilizability} by interconnecting communication link reliability with the system's unstable eigenvalues and derive a closed-form expression that quantifies this metric. 
Numerical results are provided to visualize how communication and control systems' parameters affect the probability of stabilizability of the overall system.


\end{abstract}


\section{Introduction}
\IEEEPARstart{N}{owadays}, smart grids (SGs) have expanded significantly to cover large geographical areas, including megacities, driven by the increased integration of renewable energy sources and distributed energy storage systems \cite{Micro_grid}. This rapid growth has led to a sharp rise in the number of sensors and controllers needed to monitor and manage various states and the performance of closed-loop SG systems. As the number of these components proliferates, ensuring reliable wired connections between them becomes increasingly difficult. 
In this context, wireless connectivity offers a more efficient solution to address the connectivity challenges between these components. Consequently, the closed-loop control systems of SGs can now be modeled as wireless networked control systems (WNCSs), enabling seamless wireless communication among the different elements of SGs. These WNCSs typically consist of multiple dynamic-state generators and distributed energy storage systems acting as plants, numerous sensors for monitoring, a remote controller, and actuators that execute the controller's commands on the relevant plants.
As WNCS components rely on a shared wireless medium for communication, issues like packet loss and latency significantly challenge the system's stability \cite{Review_WNCS}. Even if we assume the controller is optimally designed to generate control actions, a typical WNCS operates as a stochastic system due to the unpredictable nature of the wireless channel between the network and the controller. This randomness in the wireless channel means the plant’s behavior cannot be deterministically stabilized because of the probabilistic variations in channel quality.
One of the core challenges in WNCSs, therefore, is to enhance the probability of asymptotic stabilizability, which is directly influenced by the outage probability of the wireless link between the sensor and controller. A higher outage probability correlates with a lower probability of achieving asymptotic stabilizability. Consequently, improving the reliability of the sensor-controller link is key to increasing the chances of asymptotic stabilizability. In fact, the probability of asymptotic stabilizability can be seen as directly equivalent to the reliability of the sensor-controller link.


It is crucial to recognize that SG applications, such as load frequency control, demand ultra-high reliability for the communication link. Even small deviations in frequency from the reference value can result in severe damage and potentially cause large-scale power system failures. In this context, reliability refers to the probability of successfully transmitting plant state information within a specified transmission delay. This can be expressed as the probability that the transmission rate of the link exceeds a certain threshold set by the delay requirements.
The term ``ultra-reliability" is often used to describe the exceedingly high levels of probability needed for such applications, typically represented by several ``nines" (e.g., 99.9999\%). Given that the probability of asymptotic stabilizability in WNCSs is directly linked to the reliability of the sensor-controller link, this level of ultra-reliability can also be used to quantify the probability of asymptotic stabilizability for the system.

{\textbf{Related Works:}}
The challenge of closed-loop control under communication constraints has been widely studied in the literature. Early research on the minimum information rate required for the sensor-to-controller link to guarantee asymptotic stabilizability was presented in \cite{C_under_comm} and \cite{CO_noisy_ch}. Specifically, \cite{C_under_comm} proposed a framework for analyzing control systems with a noiseless digital communication channel between the sensor and the controller. The study showed that the transmission rate must be at least the sum of the logarithms of the unstable eigenvalues of the system’s dynamic matrix to achieve asymptotic observability and stabilizability. This framework was later extended in \cite{CO_noisy_ch} using information-theoretic principles from source and channel coding. The authors demonstrated that the rate condition from \cite{C_under_comm} applies across various channel models, including noiseless digital channels with delay, erasure channels, and memoryless Gaussian channels.

Building on these results, the authors of \cite{Rate-Cost_T} addressed the problem of minimizing a quadratic cost function related to state variables and control actions, commonly referred to as the linear quadratic regulator (LQR). In their work, the sensor and controller communicate via a wireless channel, and the authors introduced a rate-cost function to define the tradeoff between the transmission rate and the expected LQR cost. This function sets a lower bound on the minimum transmission rate necessary to achieve a given LQR cost, effectively quantifying the minimum directed mutual information required between channel input and output to meet the target cost.

Expanding on the rate-cost concept, the study in \cite{Leveraging} explored the joint optimization of LQR cost and energy consumption in centralized WNCSs with ultra-reliable and low-latency communication (URLLC) systems. The authors introduced a metric called energy-to-control efficiency (ECE), which balances spectral efficiency, LQR cost, and energy consumption, ensuring URLLC performance in IoT control systems. Similarly, \cite{R3C} investigated the joint optimization of reliability and LQR control cost in a reconfigurable intelligent surface (RIS)-assisted WNCS, introducing the reliability-to-control efficiency (RCE) metric. This work applied the rate-cost function to evaluate control performance under communication constraints and optimize factors such as transmission power, time, beamforming, and RIS reflection coefficients.
Finally, \cite{space} proposed a communication scheme for sensing-communication-computing-control (SC3) loops in deep space exploration systems. By modeling the system under the finite block length (FBL) regime, they optimized block length and transmit power to minimize the LQR cost of the SC3 loops.

{\textbf{Paper Contributions:}}
To the best of our knowledge, no previous research has established a mathematical relationship between asymptotic stabilizability and the wireless channel parameters of the sensor-to-controller link in WNCSs. This paper addresses this gap by developing such a relationship. As discussed in \cite{C_under_comm}, plant stability is determined by the eigenvalues of its dynamic matrix, a concept we will explore further. Specifically, eigenvalues with magnitudes greater than 1 indicate unstable plant behavior unless control actions are applied to stabilize it. Our main contribution is introducing the probability of asymptotic stabilizability in WNCSs and then deriving a closed-form expression that links the magnitude of the plant's unstable eigenvalues to the channel parameters. We show that as the product of a plant's unstable eigenvalues increases, the probability of asymptotic stabilizability decreases exponentially for a given set of channel parameters. In other words, a higher product of eigenvalues, indicating faster instability, requires exponential improvements in channel parameters to maintain the same probability of stabilizability.

{\textbf{Paper Structure:}}
The remainder of this paper is structured as follows: Section \ref{systemmodel} introduces the system model and notations considered in this paper. Section \ref{analysis} presents the analysis of the relationship between the sensor-controller link reliability and the probability of asymptotic stabilizability, along with the derived closed-form expression. Finally, the numerical results are discussed in Section \ref{result}, and the conclusions are provided in Section \ref{conclusion}.

\begin{figure}[!t]
\centering
\includegraphics[width=1\linewidth]{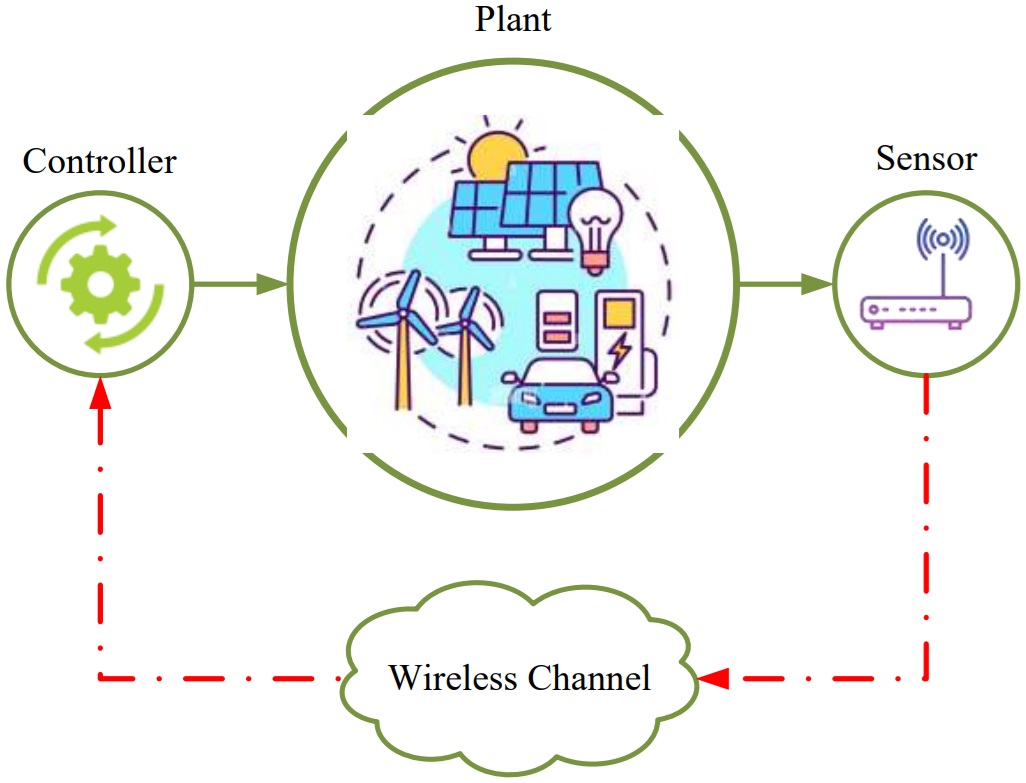}
\caption{Wireless networked control system model.}
\label{sys-model}
\end{figure}

\section{System model and Notations}\label{systemmodel}

As depicted in Fig. \ref{sys-model}, this paper considers a single loop WNCS comprising of a plant, a single sensor that monitors the plant state, and a single controller co-located with the plant, which applies the necessary control actions. The dynamic of this WNCS is modeled using the following linear time-invariant (LTI) equations \cite{C_under_comm, Rate-Cost_T} \color{black}:
\begin{equation}
\label{Dsys}
\begin{aligned}
    &\mathbf{x}[t+1] = \mathbf{A}\mathbf{x}[t] + \mathbf{B}\mathbf{u}[t] + \mathbf{w}[t], \\
    &\mathbf{y}[t] = \mathbf{C}\mathbf{x}[t],
\end{aligned}
\end{equation}
where $ t $ denotes the time step of the sampling process. Besides, $\mathbf{A} \in \mathbb{R}^{M \times M}$ and $\mathbf{B} \in \mathbb{R}^{M \times N}$ are the system and input matrices of the plant, respectively, in which $ M $ stands for the number of states and $ N $ is the number of system inputs. Also,  $\mathbf{C} \in \mathbb{R}^{1 \times M}$ is the measurement matrix. Here, $\mathbf{x}[t] \in \mathbb{R}^{M \times 1}$ represents the plant states, $\mathbf{u}[t] \in \mathbb{R}^{N \times 1}$ is the control input, $\mathbf{y}[t]$ is the sensor measurement, and $\mathbf{w}[t] \in \mathbb{R}^{N \times 1}$ denotes white Gaussian noise with zero mean and covariance matrix $\mathbf{\Sigma} \in \mathbb{R}^{M \times M}$, such that $\mathbf{w}[t] \sim \mathcal{N}(0, \mathbf{\Sigma})$. 

Additionally, it is assumed that the system is fully measurable, so measurements are not considered in this paper; only the state equation is used for further analysis. In this context, if the system described by (\ref{Dsys}) is treated as a discrete-time linear stochastic system, stabilizing the behavior of this system is a stochastic process. In this regard the asymptotic stabilizability of the system is defined in \cite{C_under_comm} as:

\textit{Definition 1}: A WNCS represented in (\ref{Dsys}) is asymptotically stabilizable if the following conditions hold for it.
\begin{enumerate}
    \item Stability: $ \forall\epsilon > 0, \exists \delta(\epsilon) $ such that $ || \mathbf{x}[0] ||_{2} \leq \delta(\epsilon) $ results in $ || \mathbf{x}[t]||_{2} \leq \epsilon, \forall t \geq 0 $.
    \item Uniform attractivity: $ \forall\epsilon > 0 $ and $ \forall\delta > 0, \exists T(\delta, \epsilon) $ such that $ || \mathbf{x}[0] ||_{2} \leq \delta $ leads to $ || \mathbf{x}[t] ||_{2} \leq \epsilon, \forall t \geq T $.
\end{enumerate}
The first condition implies that the state cannot grow unboundedly for a bounded $ \mathbf{x}[0] $ and the second one demonstrates that the state converges to zero uniformly in $ \mathbf{x}[0] $.

It has been proven in \cite{CO_noisy_ch} that the necessary condition for achieving asymptotic stabilizability is determined by the transmission rate between the sensor and the controller, denoted as $\mathcal{R}_{\text{s2c}}$ (bits per symbol). This rate must satisfy the following inequality:
\begin{equation} 
 \label{LB_rate}
 \mathcal{R}_{\text{s2c}} \geq \sum_{\left| \lambda \left( \mathbf{A}\right) \right| >1} \log_2 \left| \lambda \left( \mathbf{A}\right) \right|,
\end{equation}
where $\lambda \left( \mathbf{A}\right)$ represents the eigenvalues of the state matrix $\mathbf{A}$. This inequality indicates that the required transmission rate on the wireless link depends solely on the unstable eigenvalues of matrix $\mathbf{A}$ with magnitudes greater than $1$ ($\left| \lambda \left( \mathbf{A}\right) \right| >1$). Furthermore, it can be noted that the stable modes of the system, where $\left| \lambda \left( \mathbf{A}\right) \right| \leq 1$, can remain bounded even when the transmission rate is zero \cite{Rate-Cost_T}.

According to Shannon's information theorem, the channel capacity between the sensor and the controller is given by:
\begin{equation} 
 \label{shannon_c}
 \mathcal{C}_{\text{s2c}} = \log_2 \left( 1 + \mathrm{SNR}\right).
\end{equation}
In (\ref{shannon_c}), $\mathrm{SNR}$  is obtained from:
\begin{equation} 
 \label{snr}
 \mathrm{SNR} = \frac{|\mathcal{H}|^2 \mathcal{P}_{\text{t}}}{\mathcal{N}_0},
\end{equation}
in which \( \mathcal{H} \) represents the channel gain factor, \( \mathcal{P}_{\text{t}} \) is the transmission power of the sensor, and \( \mathcal{N}_0 \) is the power of additive white Gaussian noise (AWGN).

Based on Shannon's theorem, the transmission rate $\mathcal{R}_{\text{s2c}}$ must not exceed the channel capacity, denoted as $\mathcal{C}_{\text{s2c}}$. This condition is expressed as:
\begin{equation} 
 \label{shannon_rate}
 \mathcal{R}_{\text{s2c}} \leq \mathcal{C}_{\text{s2c}}.
\end{equation}
Therefore, by combining the conditions from (\ref{LB_rate}) and (\ref{shannon_rate}), the necessary condition for the channel capacity to ensure asymptotic stabilizability of the plant is given by:
\begin{equation} 
 \label{LB_C}
 \log_2 \left( 1 + \mathrm{SNR}\right) \geq \sum_{\left| \lambda \left( \mathbf{A}\right) \right| >1} \log_2 \left| \lambda \left( \mathbf{A}\right) \right|.
\end{equation}
This inequality ensures that the channel's SNR is high enough to achieve the required transmission rate for asymptotic stabilizability in WNCS.

In the next two definitions, we define the probability of asymptotic stabilizability and sensor-controller link reliability in detail.

\textit{Definition 2}: The necessary condition required for the channel capacity to guarantee the asymptotic stabilizability of a WNCS, as defined in \eqref{LB_C}, may not always be satisfied due to the inherent variability in wireless channel quality. In other words, the inevitable random fluctuations in the wireless channel prevent the deterministic achievement of asymptotic stabilizability. As a result, we define the probability of achieving asymptotic stabilizability in the WNCS as follows:
\begin{equation} 
 \label{stabilizability}
 \beta = \Pr \left( \log_2 \left( 1 + \mathrm{SNR}\right) \geq \sum_{\left| \lambda \left( \mathbf{A}\right) \right| >1} \log_2 \left| \lambda \left( \mathbf{A}\right) \right| \right).
\end{equation}

\textit{Definition 3}: As described in \cite{RAN_Slicing, Deep-Reliable, Statistical-Reliable}, the reliability of the sensor-controller link, denoted by $\alpha$, is defined as the probability that the transmission rate on the link meets or exceeds a specified threshold, i.e., $\alpha \left( \mathcal{R}_{th} \right) = \Pr \left( \mathcal{R}_{\text{s2c}} \geq \mathcal{R}_{th} \right)$. Given the condition in \eqref{shannon_rate}, it can be stated that $\mathcal{C}_{\text{s2c}} \geq \mathcal{R}_{\text{s2c}} \geq \mathcal{R}_{th}$. Consequently, using \eqref{shannon_c} and \eqref{shannon_rate}, the sensor-controller link reliability can be stochastically expressed as:
\begin{equation} 
 \label{reliability}
 \alpha \left( \mathcal{R}_{th} \right) = \Pr \left( \log_2 \left( 1 + \mathrm{SNR}\right) \geq \mathcal{R}_{th}\right).
\end{equation}

\section{Analysis of Reliability-Stabilizability Relationship}\label{analysis}

This section explores the relationship between the probability of asymptotic stabilizability and sensor-controller link reliability in the WNCS, aiming to derive a closed-form expression connecting them. To achieve this, Considering \eqref{stabilizability} and \eqref{reliability}, the following lemma can be easily verified. 

\textit{Lemma 1}: For WNCS, considered in \eqref{Dsys}, a desired probability of asymptotic stabilizability, denoted by $ \beta^{\text{des}} $, is feasible if and only if:  
\begin{equation} 
 \label{Trade-off}
 \alpha \left( \sum_{\left| \lambda \left( \mathbf{A}\right) \right| >1} \log_2 \left| \lambda \left( \mathbf{A}\right) \right| \right) \geq \beta^{\text{des}}.
\end{equation}
 



According to this lemma, we aim to quantify the asymptotic stabilizability of WNCS by deriving a closed-form mathematical relationship between sensor-controller link reliability and the system's unstable eigenvalues. To do this, using the property that \( \sum_{\left| \lambda \left( \mathbf{A}\right) \right| >1} \log_2 \left| \lambda \left( \mathbf{A}\right) \right| = \log_2 \prod_{\left| \lambda \left( \mathbf{A}\right) \right| >1} \left| \lambda \left( \mathbf{A}\right) \right| \), we first reformulate the sensor-controller link reliability in (\ref{Trade-off}) as follows:
\begin{equation} 
 \label{1st-step}
 \alpha = \Pr \left( \log_2 \left( 1 + \mathrm{SNR} \right) \geq \log_2 \prod_{\left| \lambda \left( \mathbf{A}\right) \right| >1} \left| \lambda \left( \mathbf{A}\right) \right| \right).
\end{equation}
By canceling the logarithms on both sides of the inequality in (\ref{1st-step}), the sensor-controller link reliability simplifies to:
\begin{equation} 
 \label{2nd-step}
 \alpha = \Pr \left( \mathrm{SNR} \geq \left( \prod_{|\lambda \left( \mathbf{A}\right)| > 1} \left| \lambda \left( \mathbf{A}\right) \right| \right) - 1 \right).
\end{equation}

To derive a closed-form expression for the sensor-controller link reliability \( \alpha \), we must first determine the probability distribution of \( \mathrm{SNR} \) in (\ref{2nd-step}). From the SNR equation in (\ref{snr}), the only stochastic component is the channel gain factor \( \mathcal{H} \). We model the channel as a Rayleigh flat fading channel with path loss, as described in \cite{WBAN}:
\begin{equation} 
 \label{loss}
 \mathcal{L}_{\text{d}} = \frac{\mathcal{P}_{\text{t}}}{\mathcal{P}_{\text{r}}} = (d_{\text{s2c}})^\eta \mathcal{L}_0,
\end{equation}
where \( \mathcal{P}_{\text{r}} \) is the received power, \( d_{\text{s2c}} \) is the distance between the sensor and controller, \( \eta \) is the path loss exponent, and \( \mathcal{L}_0 \) is the path loss at the reference distance \( d_0 = 1 \text{m} \).

Moreover, due to the presence of dense scattering objects, the magnitude of the fading channel coefficient \( |h| \) follows a Rayleigh distribution with a probability density function \( P(|h|) = \frac{2|h|}{\Omega} \exp \left(-\frac{|h|^2}{\Omega} \right) \), where \( \Omega = \mathbb{E}[ |h|^2 ] \). 

By combining the path loss from \eqref{loss} with the fading channel coefficient \( |h| \), the instantaneous channel gain \( \mathcal{H} \) is expressed as \cite{WBAN}:
\begin{equation} 
 \label{ch-gain}
 \mathcal{H} = \frac{h}{\sqrt{\mathcal{L}_{\text{d}}}} = \frac{h}{\sqrt{(d_{\text{s2c}})^\eta \mathcal{L}_0}}.
\end{equation}

Substituting (\ref{snr}) and (\ref{ch-gain}) into (\ref{2nd-step}), the sensor-controller link reliability \( \alpha \) is expressed as:
\begin{equation} 
 \label{reform_alpha}
 \begin{aligned}
    \alpha 
    &= 
    \Pr \left( 
        \frac{|h|^2 \mathcal{P}_{\text{t}}}{\mathcal{N}_0 \mathcal{L}_0 (d_{\text{s2c}})^\eta} \geq \left( \prod_{|\lambda \left( \mathbf{A}\right)| > 1} \left| \lambda \left( \mathbf{A}\right) \right| \right) - 1 
    \right) 
    \\
    &=
    \Pr \left(
        |h| \geq \sqrt{\frac{\mathcal{N}_0 \mathcal{L}_0 (d_{\text{s2c}})^\eta}{\mathcal{P}_{\text{t}}} \left( \left( \prod_{|\lambda \left( \mathbf{A}\right)| > 1} \left| \lambda \left( \mathbf{A}\right) \right| \right) - 1 \right)} 
    \right).
 \end{aligned}
\end{equation}
Since \( |h| \) follows a Rayleigh distribution, the sensor-controller link reliability \( \alpha \) corresponds to the Complementary Cumulative Distribution Function (CCDF) of the Rayleigh distribution \cite{Reliability}, given by:
\begin{equation} 
 \label{final_reliability}
 \begin{aligned}
 \alpha &= \mathrm{CCDF} \left( \sqrt{\frac{\mathcal{N}_0 \mathcal{L}_0 (d_{\text{s2c}})^\eta}{\mathcal{P}_{\text{t}}} \left( \left( \prod_{|\lambda \left( \mathbf{A}\right)| > 1} \left| \lambda \left( \mathbf{A}\right) \right| \right) - 1 \right)} \right) \\
 &= \exp \left( -\frac{\mathcal{N}_0 \mathcal{L}_0 (d_{\text{s2c}})^\eta \left( \left( \prod_{|\lambda \left( \mathbf{A}\right)| > 1} \left| \lambda \left( \mathbf{A}\right) \right| \right) - 1 \right)}{2 \mathcal{P}_{\text{t}}} \right).
 \end{aligned}
\end{equation}
Taking the natural logarithm of both sides of equation (15) and rearranging yields a closed-form expression as:
\begin{equation} 
 \label{closed-form}
\prod_{|\lambda\left( \mathbf{A}\right)| >1} \left| \lambda \left( \mathbf{A}\right) \right| = -\frac{2 \mathcal{P}_{\text{t}}}{\mathcal{N}_0 \mathcal{L}_0 (d_{\text{s2c}})^\eta} \ln \alpha + 1.
\end{equation}


Equation \eqref{closed-form} establishes a clear mathematical relationship between the product of unstable eigenvalues of the WNCS and the reliability of the sensor-controller link. Specifically, for a WNCS with a defined product of unstable eigenvalues, represented as \( \prod_{|\lambda \left( \mathbf{A}\right)| \geq 1} \left| \lambda \left( \mathbf{A}\right) \right| \), the maximum achievable sensor-controller link reliability equals \( \alpha \) given certain channel parameters, including \( \mathcal{N}_0 \), \( \mathcal{L}_0 \), \( \eta \), \( d_{\text{s2c}} \), and \( \mathcal{P}_{\text{t}} \). Furthermore, according to \textit{Lemma 1}, the sensor-controller link reliability is directly equivalent to the probability of asymptotic stabilizability \( \beta \). Thus, equation \eqref{closed-form} shows that, for the specified WNCS, the maximum achievable probability of asymptotic stabilizability is exactly \( \alpha \). This relationship can be accurately assessed using this closed-form expression.

It is worth noting that in WNCS with a high product of unstable eigenvalues, we can omit  \( 1 \) from the right-hand side of \eqref{closed-form}. In this regard, if the product of unstable eigenvalues doubles, the sensor-controller link reliability decreases by a factor of \( \exp(2) \). According to  \textit{Lemma 1}, this reduction happens because, in a WNCS with a doubled product of unstable eigenvalues, \( \mathcal{R}_{th} \) increases by \( \log_2(2) = 1 \).

Furthermore, in what follows, we consider two general cases of a WNCS with $K$ loops. In both cases, all sensors are assumed to transmit data to their respective controllers over the same frequency band and at equal power levels. We then investigate how interference from these concurrent transmissions affects the reliability of the sensor-controller links. In this regard, $\mathcal{N}_0$ is assumed to be negligible compared to the interference.

\subsubsection{Single Interference Case}
Given $K=2$, the signal-to-interference ratio (SIR) for the $i$-th link is given by $\mathrm{SIR}_{i} = \frac{h_i}{h_j}\left( \frac{d_i}{d_j}\right)^{-\eta}, \quad i,j \in \{1,2\}, \ i \neq j$, where $h_i$ represents the Rayleigh fading channel coefficient, and $d_i$ is the Euclidean distance between sensor $i$ and its corresponding controller. The reliability of the $i$-th sensor-controller link is calculated as:
\begin{equation}
 \label{S-inter}
 \alpha_{i} = \Pr \left[ \frac{h_i}{h_j} \geq \left( \prod_{|\lambda \left( \mathbf{A}\right)| > 1} \left| \lambda \left( \mathbf{A}\right) \right| \right)\left( \frac{d_j}{d_i}\right)^{-\eta} - \left( \frac{d_j}{d_i}\right)^{-\eta} \right].
\end{equation}
According to CCDF, introduced in \cite{Meta}, after some straightforward mathematical calculations $ \alpha_{i} $ is achieved as:
\begin{equation}
 \label{S-inter2}
 \alpha_{i} = \frac{1}{1+\left( \prod_{|\lambda \left( \mathbf{A}\right)| > 1} \left| \lambda \left( \mathbf{A}\right) \right| \right)\left( \frac{d_j}{d_i}\right)^{-\eta} - \left( \frac{d_j}{d_i}\right)^{-\eta}}.
\end{equation}
\subsubsection{Full Interference Case}
Assuming that simultaneous transmissions from all $K>2$ loops interfere with each other, the SIR for the $i$-th link is expressed as $\mathrm{SIR}_{i} = \frac{h_{i}(d_{i})^{-\eta}}{\sum_{j=1, j\neq i}^{K}h_{j}(d_{j})^{-\eta}}$ \cite{Meta}.
Thus, the reliability of the $i$-th sensor-controller link is given by:
\begin{equation}
 \label{F-inter}
 \alpha_{i} = \Pr \left[ \frac{h_{i}(d_{i})^{-\eta}}{\sum_{j=1, j\neq i}^{K}h_{j}(d_{j})^{-\eta}} \geq \left( \prod_{|\lambda \left( \mathbf{A}\right)| > 1} \left| \lambda \left( \mathbf{A}\right) \right| \right) - 1 \right].
\end{equation}
Consequently, the CCDF of \eqref{F-inter} is obtained from:
\begin{equation}
 \label{F-inter2}
 \alpha_{i} = \frac{(d_{i})^{-\eta}}{(d_{i})^{-\eta} + \sum_{j=1, j\neq i}^{K}(d_{j})^{-\eta} \left[ \left( \prod_{|\lambda \left( \mathbf{A}\right)| > 1} \left| \lambda \left( \mathbf{A}\right) \right| \right) - 1 \right]}.
\end{equation}

\section{Numerical Results}\label{result}
This section explores how changes in parameters such as transmission power, distance, and path loss exponent impact sensor-controller link reliability, which is associated with the probability of asymptotic stabilizability. To conduct a thorough evaluation of the closed-form expression in (\ref{closed-form}), we analyze a diverse set of closed-loop control applications characterized by different products of unstable eigenvalues. To achieve this, we perform an extensive search across various use cases of WNCSs to extract their dynamic matrix \( \mathbf{A} \). Next, we calculate the eigenvalues of \( \mathbf{A} \) and derive the product of unstable eigenvalues as well as the transmission rate threshold for each use case, as illustrated in Table \ref{tab:my_label}. As highlighted in Table \ref{tab:my_label}, there are significant variations in the product of unstable eigenvalues among different use cases. Consequently, in simulation results, a wide range of products of unstable eigenvalues is included to address various practical applications of WNCSs.

\begin{table}
    \centering
    \caption{the product of unstable eigenvalues and transmission rate threshold for different use-cases of WNCSs}
    \label{tab:my_label}
    \adjustbox{width=\columnwidth}{
    \begin{tabular}{|c|c|c|}
         \hline
        \textbf{Use-case} & $\prod_{|\lambda\left( \mathbf{A}\right)| >1} \left| \lambda \left( \mathbf{A}\right) \right|$ & $\mathcal{R}_{th}$ \\
        \hline
        Voltage Regulation in DC Micro Grids \cite{Voltage-R} & $6\times10^{7}$ & $25.83$\\
        \hline
        Load Frequency Control \cite{Event_Trig} & $412.99$ & $8.69$ \\        
        \hline
        Adaptive Cruise control \cite{cruise-c}& 2.2 & $1.1375$\\
        \hline
    \end{tabular}
    }
\end{table}

In what follows, we consider three different scenarios to evaluate the closed-form expression. For all scenarios, the path loss at the reference distance is set to \( \mathcal{L}_{0} = 0.1 \) mW, and the AWGN power is set to \( \mathcal{N}_{0} = 10 \) mW.

\begin{figure}[!t]
\centering
\includegraphics[scale=0.6]{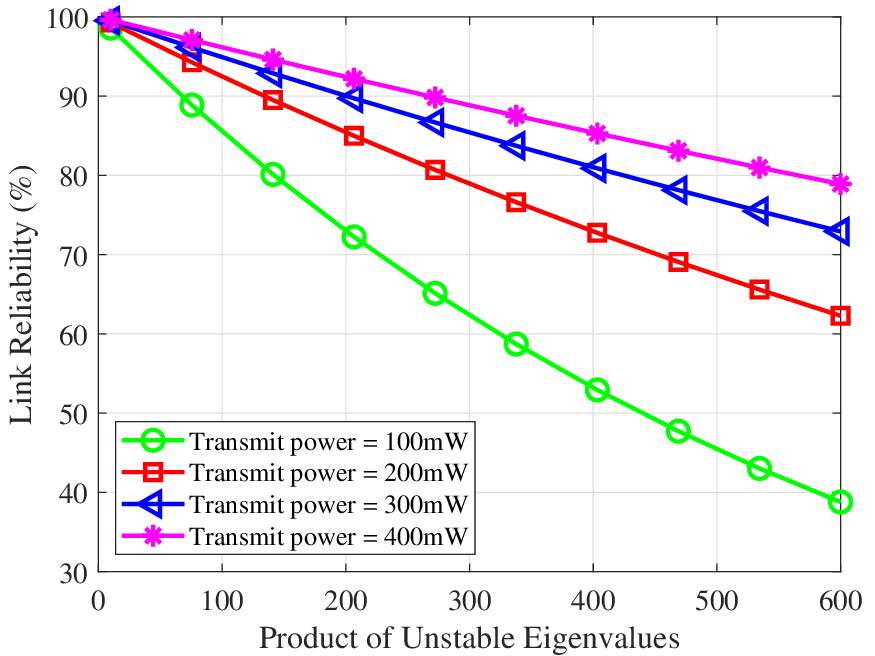}
\caption{Variation of sensor-controller link reliability with respect to different products of unstable eigenvalues in WNCS, considering various transmission power levels of the sensor.}
\label{power_var}
\end{figure}

\textbf{Scenario I:} In this scenario, we investigate how sensor-controller link reliability changes with different closed-loop system dynamics at various transmission power levels, assuming \( d = 10 \) m and \( \eta = 2.5 \). The sensor's transmission power ranges from \( \mathcal{P}_{\text{t}} = 100 \) mW to \( \mathcal{P}_{\text{t}} = 400 \) mW. As illustrated in Fig. \ref{power_var}, increasing the product of the eigenvalues of the closed-loop system from 10 to 600 results in a consistent decrease in sensor-controller link reliability. A higher product of eigenvalues signifies greater variability in system states in response to minor changes. Therefore, Fig. \ref{power_var} indicates that to maintain high sensor-controller link reliability in a WNCS with rapid dynamics, a significant increase in transmission power is necessary. For example, in a typical load frequency control application where the product of eigenvalues is 600 \cite{Event_Trig}, a transmission power of 100 mW yields a sensor-controller link reliability of only 40 percent. To achieve an 80 percent reliability, the transmission power must be increased fourfold. In other words, to boost the probability of asymptotic stabilizability from 40 percent to 80 percent, the transmission power from the sensor to the controller needs to be fourfold. 
Furthermore, it is evident that with a fixed transmission power, such as \( \mathcal{P}_{\text{t}} = 100 \) mW, increasing the product of unstable eigenvalues from 200 to 400 causes the sensor-controller link reliability, or the probability of asymptotic stabilizability, to drop from 72 percent to 53 percent. This decline results from the increase in the lower bound of \( \mathcal{R}_{th} \) from \( \log_2(200) = 7.6439 \) to \( \log_2(400) = 8.6439 \).

\begin{figure}[!t]
\centering
\includegraphics[scale=0.6]{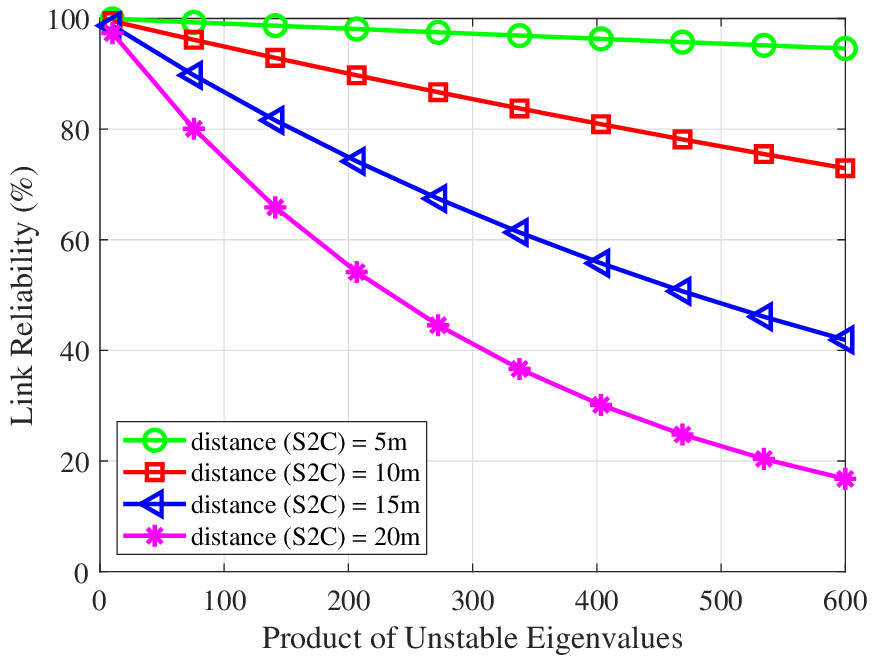}
\caption{Variation of sensor-controller link reliability versus different products of unstable eigenvalues in WNCSs, according to the sensor to controller distance.}
\label{dis_var}
\end{figure}

\textbf{Scenario II:} In this scenario, we examine how varying the transmission distance from the sensor to the controller affects link reliability across different applications characterized by varying products of unstable eigenvalues. We assume that the transmission power and path loss exponent remain constant at \( \mathcal{P}_{\text{t}} = 300 \) mW and \( \eta = 2.5 \), respectively. As shown in Fig. \ref{dis_var}, increasing the transmission distance leads to a lower maximum achievable probability of asymptotic stabilizability, indicating a decline in sensor-controller link reliability. Moreover, the influence of the product of unstable eigenvalues on link reliability becomes more significant at greater distances. This is due to the distance being raised to the power of the path loss exponent (\( \eta = 2.5 \)) in (\ref{closed-form}), which amplifies its impact on sensor-controller link reliability. Importantly, due to the exponential relationship between link reliability, transmission distance, and the product of unstable eigenvalues, achieving higher sensor-controller link reliability in WNCSs with a high product of unstable eigenvalues requires a much greater reduction in transmission distance compared to those with a lower product of unstable eigenvalues.

\begin{figure}[!t]
\centering
\includegraphics[scale=0.6]{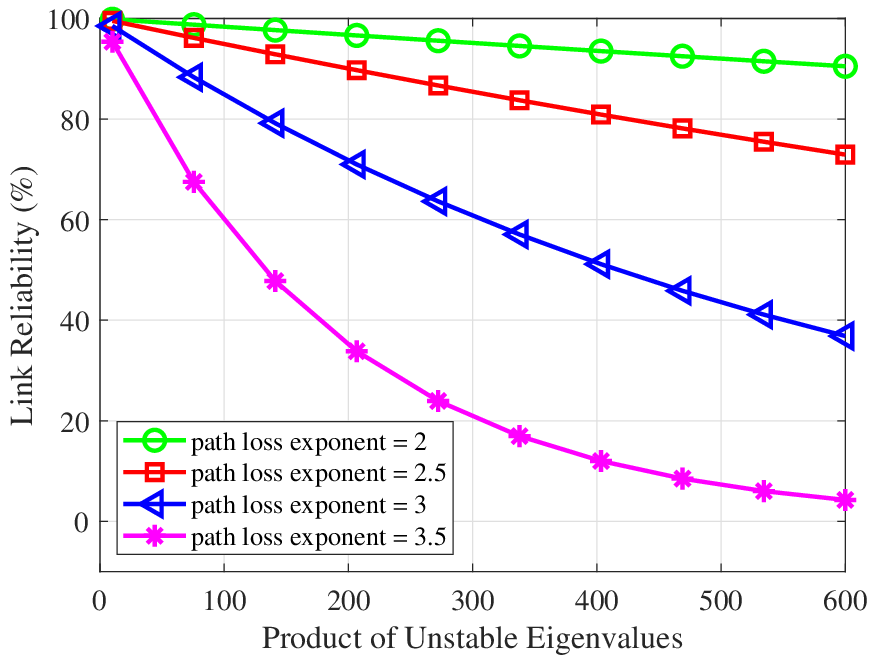}
\caption{Variation of sensor-controller link reliability versus different products of unstable eigenvalues in WNCSs, according to sensor's path loss exponents.}
\label{etha_var}
\end{figure}

\textbf{Scenario III:} 
This scenario evaluates how different propagation environments influence the probability of asymptotic stabilizability, assuming fixed transmission power and distance at \( \mathcal{P}_{\text{t}} = 300 \) mW and \( d = 10 \) m, respectively. Variations in environmental characteristics are represented by different path loss exponents. As illustrated in Fig. \ref{etha_var}, the decline in sensor-controller link reliability due to an increase in the product of unstable eigenvalues is much more pronounced in environments with higher path loss exponents, such as \( \eta = 3.5 \). This is attributed to the increased density of obstacles with varying absorption properties, leading to greater signal attenuation. In essence, a WNCS with a specific product of unstable eigenvalues is likely to achieve a lower probability of asymptotic stabilizability in environments with dense obstacles compared to those that are sparser. For instance, Fig. \ref{etha_var} shows that a load frequency control application in a smart grid, with a product of unstable eigenvalues around \( 600 \), can achieve a maximum sensor-controller link reliability of \( 39 \) percent (equivalent to a probability of asymptotic stabilizability) in an environment with \( \eta = 3 \). However, in an environment with \( \eta = 3.5 \), the sensor-controller link reliability significantly drops to \( 2 \) percent.

\section{Conclusion}\label{conclusion}
This paper addressed the challenge of analyzing the probability of asymptotic stabilizability in WNCSs that use wireless communication to connect the sensor to the controller, considering various communication impairments. First, it was demonstrated that sensor-controller link reliability is equivalent to the probability of asymptotic stabilizability in WNCSs. A closed-form mathematical relationship was then derived, linking the system's unstable eigenvalues, wireless channel characteristics, and sensor-controller link reliability. The analysis revealed that sensor-controller link reliability decreases exponentially as the product of the system's unstable eigenvalues increases. The numerical results showed that achieving a high probability of asymptotic stabilizability requires an exponential improvement in the quality of the sensor-to-controller wireless channel. These findings highlight the necessity of deploying 6G key enabling technologies for critical control applications with large unstable eigenvalues, such as load frequency control systems, where ultra-high probabilities of asymptotic stabilizability are essential to ensure a stable and reliable power supply for consumers. For instance, reconfigurable intelligent surfaces can enhance the channel quality between sensors and controllers by dynamically adjusting the propagation environment. By effectively modifying the propagation environment, these technologies can substantially increase the likelihood of asymptotic stabilizability.

\bibliographystyle{IEEEtran}

\bibliography{Refs}

\begin{thebibliography}{10}
\providecommand{\url}[1]{#1}
\csname url@samestyle\endcsname
\providecommand{\newblock}{\relax}
\providecommand{\bibinfo}[2]{#2}
\providecommand{\BIBentrySTDinterwordspacing}{\spaceskip=0pt\relax}
\providecommand{\BIBentryALTinterwordstretchfactor}{4}
\providecommand{\BIBentryALTinterwordspacing}{\spaceskip=\fontdimen2\font plus
\BIBentryALTinterwordstretchfactor\fontdimen3\font minus
  \fontdimen4\font\relax}
\providecommand{\BIBforeignlanguage}[2]{{%
\expandafter\ifx\csname l@#1\endcsname\relax
\typeout{** WARNING: IEEEtran.bst: No hyphenation pattern has been}%
\typeout{** loaded for the language `#1'. Using the pattern for}%
\typeout{** the default language instead.}%
\else
\language=\csname l@#1\endcsname
\fi
#2}}
\providecommand{\BIBdecl}{\relax}
\BIBdecl

\bibitem{Micro_grid}
Y.~Chen, C.~Li, D.~Qi, Z.~Li, Z.~Wang, and J.~Zhang, ``Distributed
  event-triggered secondary control for islanded microgrids with proper trigger
  condition checking period,'' \emph{IEEE Transactions on Smart Grid}, vol.~13,
  no.~2, pp. 837--848, Mar. 2022.

\bibitem{Review_WNCS}
Y.~Wang, S.~Wu, C.~Lei, J.~Jiao, and Q.~Zhang, ``A review on wireless networked
  control system: The communication perspective,'' \emph{IEEE Internet of
  Things Journal}, vol.~11, no.~5, pp. 7499--7524, Mar. 2024.

\bibitem{C_under_comm}
S.~Tatikonda and S.~Mitter, ``Control under communication constraints,''
  \emph{IEEE Transactions on Automatic Control}, vol.~49, no.~7, pp.
  1056--1068, July 2004.

\bibitem{CO_noisy_ch}
------, ``Control over noisy channels,'' \emph{IEEE Transactions on Automatic
  Control}, vol.~49, no.~7, pp. 1196--1201, July 2004.

\bibitem{Rate-Cost_T}
V.~Kostina and B.~Hassibi, ``Rate-cost tradeoffs in control,'' \emph{IEEE
  Transactions on Automatic Control}, vol.~64, no.~11, pp. 4525--4540, Nov.
  2019.

\bibitem{Leveraging}
H.~Yang, K.~Zhang, K.~Zheng, and Y.~Qian, ``Leveraging linear quadratic
  regulator cost and energy consumption for ultrareliable and low-latency iot
  control systems,'' \emph{IEEE Internet of Things Journal}, vol.~7, no.~9, pp.
  8356--8371, Sep. 2020.

\bibitem{R3C}
S.~Han, L.~Jin, X.~Xu, X.~Tao, and P.~Zhang, ``R3c: Reliability and control
  cost co-aware in ris-assisted wireless control systems for iiot,'' \emph{IEEE
  Internet of Things Journal}, vol.~11, no.~8, pp. 13\,692--13\,707, Apr. 2024.

\bibitem{space}
\BIBentryALTinterwordspacing
X.~Fang, W.~Feng, Y.~Chen, N.~Ge, and G.~Zheng, ``Control-oriented deep space
  communications for unmanned space exploration,'' \emph{IEEE Transactions on
  Wireless Communications}, pp. 1--1, 2024. [Online]. Available:
  \url{10.1109/TWC.2024.3414854}
\BIBentrySTDinterwordspacing

\bibitem{RAN_Slicing}
P.~Yang, X.~Xi, T.~Q.~S. Quek, J.~Chen, X.~Cao, and D.~Wu, ``Ran slicing for
  massive iot and bursty urllc service multiplexing: Analysis and
  optimization,'' \emph{IEEE Internet of Things Journal}, vol.~8, no.~18, pp.
  14\,258--14\,275, Sep. 2021.

\bibitem{Deep-Reliable}
H.~Yang, Z.~Xiong, J.~Zhao, D.~Niyato, C.~Yuen, and R.~Deng, ``Deep
  reinforcement learning based massive access management for ultra-reliable
  low-latency communications,'' \emph{IEEE Transactions on Wireless
  Communications}, vol.~20, no.~5, pp. 2977--2990, May 2021.

\bibitem{Statistical-Reliable}
M.~Angjelichinoski, K.~F. Trillingsgaard, and P.~Popovski, ``A statistical
  learning approach to ultra-reliable low latency communication,'' \emph{IEEE
  Transactions on Communications}, vol.~67, no.~7, pp. 5153--5166, July 2019.

\bibitem{WBAN}
Z.~Askari, J.~Abouei, M.~Jaseemuddin, and A.~Anpalagan, ``Energy-efficient and
  real-time noma scheduling in iomt-based three-tier wbans,'' \emph{IEEE
  Internet of Things Journal}, vol.~8, no.~18, pp. 13\,975--13\,990, Sep. 2021.

\bibitem{Reliability}
A.~Gomes, J.~Kibiłda, and L.~A. DaSilva, ``Assessing the spectrum needs for
  network-wide ultra-reliable communication with meta distributions,''
  \emph{IEEE Communications Letters}, vol.~27, no.~8, pp. 2242--2246, Aug.
  2023.

\bibitem{Meta}
M.~Haenggi, ``Meta distributions—part 2: Properties and interpretations,''
  \emph{IEEE Communications Letters}, vol.~25, no.~7, pp. 2094--2098, July
  2021.

\bibitem{Voltage-R}
Y.~Yu, G.-P. Liu, and W.~Hu, ``Blockchain protocol-based secondary predictive
  secure control for voltage restoration and current sharing of dc
  microgrids,'' \emph{IEEE Transactions on Smart Grid}, vol.~14, no.~3, pp.
  1763--1776, May 2023.

\bibitem{Event_Trig}
X.~Zhao, S.~Zou, P.~Wang, and Z.~Ma, ``Bandwidth-aware event-triggered load
  frequency control for power systems under time-varying delays,'' \emph{IEEE
  Transactions on Power Systems}, vol.~38, no.~5, pp. 4530--4541, Sep. 2023.

\bibitem{cruise-c}
\BIBentryALTinterwordspacing
J.~Zhao, Z.~Wang, Y.~Lv, J.~Na, C.~Liu, and Z.~Zhao, ``Data-driven learning for
  h$_{\infty }$ control of adaptive cruise control systems,'' \emph{IEEE
  Transactions on Vehicular Technology}, pp. 1--15, 2024. [Online]. Available:
  \url{10.1109/TVT.2024.3447060}
\BIBentrySTDinterwordspacing

\end{thebibliography}
 
\vfill

\end{document}